\documentstyle{jaa}

\input epsf.sty
\begin{document}
\title[ Radiation transfer in H II regions]
{A new scheme of radiation transfer in H II regions  \\
          including transient heating of grains}
\author[S.K. Ghosh \& R.P. Verma]
{S. K. Ghosh\thanks{E-mail : swarna@tifr.res.in} ~~\&~~ R. P. Verma \\
Tata Institute of Fundamental Research, 
Homi Bhabha Road, Bombay 400 005 } 
\pubyear{2000}
\volume{21}
\pagerange{\pageref{firstpage}--\pageref{lastpage}}
\setcounter{page}{61}
\date{Received 2000 April 15; accepted 2000 May 11}
\maketitle
\label{firstpage}

\begin{abstract}
 A new scheme of radiation transfer for understanding infrared spectra 
of H II regions, has been developed. This scheme considers  
non-equilibrium processes (e.g. 
 transient heating of the very small grains, VSG;
and the polycyclic aromatic hydrocarbon, PAH)
also, in addition to the equilibrium thermal emission from normal
dust grains (BG).
The spherically symmetric interstellar dust cloud is segmented into a
large number of ``onion skin" shells in order to implement the 
non-equilibrium processes.
The scheme attempts to fit the observed SED originating from the dust
component, by exploring the following parameters :
(i) geometrical details of the dust cloud,
(ii) PAH size and  abundance,
(iii) composition of normal grains (BG),
(iv) radial distribution of all dust (BG, VSG \& PAH). 

  The scheme has been applied to a set of five compact H II regions 
(IRAS 18116--1646, 18162--2048, 19442+2427, 22308+5812, 
\& 18434--0242)
whose spectra are available with adequate spectral resolution.
The best fit models and inferences about the parameters for these 
sources are presented.

\end{abstract}

\section{Introduction}

  Till recently, the mid to far infrared 
spectral energy distribution (SED) of
Galactic star forming regions in general was
available only in the four IRAS bands (12, 25, 60 \& 100 $\mu$m).
In some relatively rare cases, spectroscopy in the 10 $\mu$m 
band through the atmospheric window, was also available.
However, the situation has changed drastically recently,
due to the advent of the Infrared Space Observatory (ISO). The
ISOPHOT photometer along with ISO-SWS and ISO-LWS spectrometers
together has revolutionized the availability of information
about SED of the astrophysical sources in general.

 In the literature, several radiation transfer schemes 
have been used for interstellar dust clouds with embedded YSOs
in spherical 
(e.g. Scoville \& Kwan 1976, Leung 1976, 
Churchwell, Wolfire, \& Wood 1990)
as well as cylindrical (Ghosh \& Tandon 1985, Dent 1988, Karnik \&
Ghosh 1999) geometries. All of these considered the dust
grains to be in thermal equilibrium.
The role of non-equilibrium processes (resulting in
transient heating / excitation of grains, particularly in the
vicinity of a source of UV radiation) has become evident from
significant near \& mid infrared continuum emission detected in 
Galactic star forming regions
(Sellgren 1984, Puget, Leger, \& Boulanger 1985, 
Boulanger, Baud \& van Albada 1985) 
as well as spectral features (Leger \& Puget 1984, 
Allamandola, Tielens, \& Barker, 1985, Puget \& Leger 1989).
The importance of these processes has also been demonstrated
in extragalactic nuclei / star forming regions 
(Moorwood {\it et al.}, 1996, Metcalfe {\it et al.}, 1996, 
Ghosh, Drapatz, \& Peppel, 1986).
A comparison of the observed mid-IR spectral features with
laboratory data, has led to the identification of a new
constituent of the interstellar medium - 
polycyclic aromatic hydrocarbons (PAH).
The enhanced continuum emission in the near \& mid IR has been mainly
attributed to the very small grains (VSG) of radii 10-100 \AA.
Hence, it is obviously important to include the non-equilibrium
processes in attempting to model the observed 
SED of star forming regions in general.
Basically, grains of very small size or a large organic molecule,
with effective heat capacity comparable
to the energy of a single UV photon get excited 
(for a short time) to an
energy state well above its thermal equilibrium state corresponding
to the local radiation field. The photons emitted during the 
de-excitation process contribute to the near / mid IR part of the
SED, which shows continuum excess and emission features which
are unexplained by radiative transfer models considering the
emission from large grains in thermal equilibrium alone.
Recently, Siebenmorgen \& Krugel (1992) have 
attempted to quantify the properties of the dust components
relevant for non-equilibrium processes (VSG \& PAH), from the
infrared data of sources in different astronomical environments
in our Galaxy.
The role of VSG on the infrared emission from 
externally heated dust clouds has been studied by Lis \& Leung (1991).
Krugel \& Siebenmorgen (1994) have presented a method to model the
transfer of radiation in dusty galactic nuclei, which includes
the presence of VSG and PAH.

  Here we present a scheme of 
radiative transfer developed by us which is applicable
in spherical geometry. This includes, 
in addition to the dust grains in thermal equilibrium (of
normal size, hereafter big grains or BG),
the transient heating of very small grains (VSG)
as well as the PAH molecules.
An attempt has been made to model
five compact H II regions : 
 IRAS 18116--1646, 18162--2048, 
 19442+2427, 22308+5812 \& 18434--0242
using the above scheme.

   In section 2, the radiative transfer modelling scheme is briefly
described. The results of modelling 
the five compact H II regions are presented in section 3. The last
section (4) consists of discussion.

\section{The Modelling Scheme }

\subsection{Dust components and their properties}

 The normal grains (BG) consist of
 two components : astronomical silicate and graphite. Their
 size distribution is taken as per Mathis, Rumpl \& Nordsieck (1977) to be
 a power law, $n(a) \propto a^{\gamma}$, with --3.5 as the exponent. 
The lower and upper limits
 of the grain radii are taken to be 0.01 $\mu$m and 0.25 $\mu$m
 respectively as recommended by Mathis, Mezger \& Panagia (1983), for 
 both astronomical silicate as well as graphite grains. 
The scattering and absorption
 coefficients, and anisotropic scattering factors have been
 taken from Draine \& Lee (1984) and Laor \& Draine (1993). 
     
     The VSG component is taken to be graphite grains of 
 a single size : either 10 \AA \  or 50 \AA \  in radius. Their
 optical properties have also been taken from Draine 
 \& Lee (1984). Abundance of VSG is connected to that of
 the normal grains through a scaling factor $Y_{VSG}$, 
 which gives the fraction of dust mass in VSG form to the
 normal BG form. 
The value of $Y_{VSG}$ was taken from Desert, Boulanger, \& Puget
(1990), which is needed to account for the 2200 \AA \ bump in the
average interstellar medium in the Galaxy,
and it has been held fixed for all models considered here.

     The PAH component is assumed to be either a single
 molecule with about 15 -- 30 atoms, or a large 
complex consisting of 10 -- 20 of these molecules
as used by Siebenmorgen (1993).
 Their optical properties, feature centres, feature shape
 and widths have been taken from Leger \& d'Hendecourt (1987).
 The abundance of PAH component is also connected through
 a scaling factor $Y_{PAH}$ to the normal grains (BG).
 There are two additional parameters explored in the modelling
 of the observed PAH spectral features :
 (i) the radius of the PAH molecule / complex, $a_{PAH}$; and
 (ii) the de-hydrogenation factor, $f_{de-H}$.
The value of $f_{de-H}$ lies between 0 and 1
($f_{de-H}$ = 0 refers to completely hydrogenated PAH).
 Whereas $a_{PAH}$ has implications of heat capacity and hence
 the efficiency of transient heating for a given radiation
 field, the $f_{de-H}$ affects the ratios of PAH features
 resulting from the C-H versus C=C stretch modes.

\subsection{Geometry}

  The star forming region is considered as a spherical
dust cloud immersed in an isotropic interstellar radiation field,
with an embedded source of energy (e.g. a ZAMS star) at its centre.
A central cavity in this cloud represents sublimation / destruction
of grains in the intense radiation field of the central source.
A schematic of the dust cloud is presented in Figure 1.

\begin {figure*}
 \epsfxsize=350.0pt
 \epsfbox {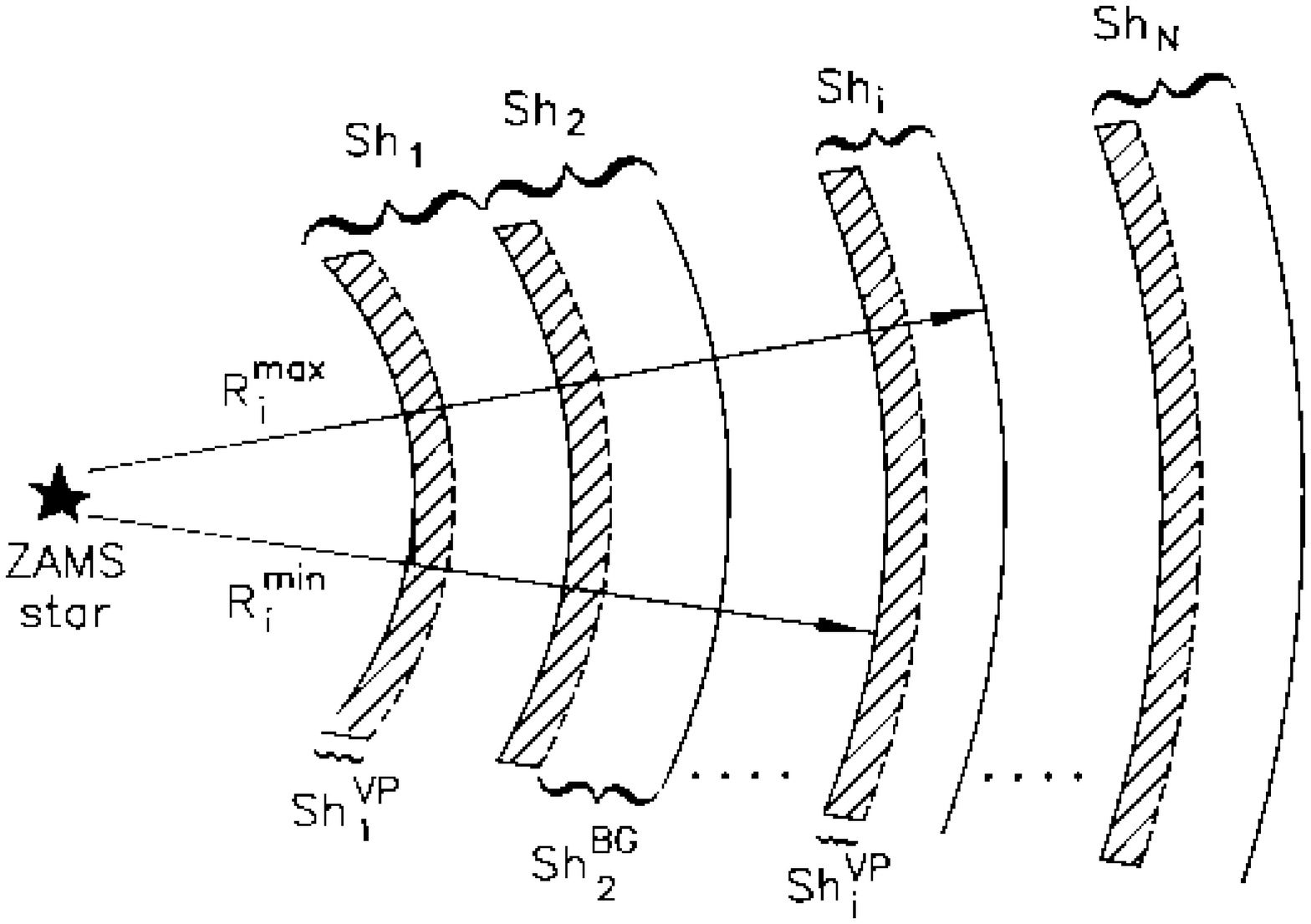}
\caption{
 Schematic diagram of the shell structure of the cloud.
}
\end {figure*}

  This spherically symmetric dust cloud, 
is divided into a large
 number of concentric contiguous spherical shells (say $Sh_{1},
 Sh_{2}, ..., Sh_{N}$) like ``onion skins". Each shell, $Sh_{i}$, 
 is identified 
 by its inner and outer radii ($R_{i}^{min} ~~ \&  ~~ R_{i}^{max}$ ;
 see Figure 1). These shells 
can be of different selectable thicknesses,
depending on the optical depth at the shortest relevant wavelength.
In order to incorporate the presence
 of both -- normal grains (BG, responsible for emission
 at thermal equilibrium), as well as the grains responsible
 for non-equilibrium emission (VSG and
  PAH), each shell is
 subdivided into a pair of sub-shells, $Sh^{BG}_{i} ~~ \& ~~
 Sh^{VP}_{i}$ corresponding to these two components 
 respectively. Whereas the former consists of only BG,
 the latter consists of only the VSG and PAH.
 
    The full detailed radiative transfer calculations 
 assuming the normal grains to be in thermal equilibrium, are 
 performed in each of the sub-shells $Sh^{BG}_{i}$, 
 for $i=1,2, ..., N$. The subshells $Sh^{VP}_{i}$, go through
 a statistical mechanical treatment describing the 
 non-equilibrium emission processes for the VSGs and the PAHs.
 For simplicity of computations,
 the sub-shells $Sh^{VP}_{i}$ are considered to be 
 very thin compared to 
the total thickness of the shell $Sh_{i}$,
and this sub-shell
 is assumed to be placed at the inner edge of the shell
 $Sh_{i}$ (see Figure 1). The final results are expected to be
 insensitive to the above simplification since individual
 shells are optically thin.
 
    Radiative transport at each of the two sub-shells is
 carried out as a two point boundary value problem, the two
 boundary conditions being the incident radiation fields at
 the two surfaces. The calculations begin with the given
 spectrum emitted by the embedded energy source (in general,
 an Initial Mass Function weighted synthetic stellar spectrum
 ensemble) incident at the inner boundary 
 of the first shell $Sh_{1}$.
 The outer surface of the last (outermost) shell, $Sh_{N}$,
 has the interstellar radiation field (ISRF) incident on it
from the outside.
Starting from the ``core" side of the
 first shell, the radiation is transported through 
 the sub-shell $Sh^{VP}_{1}$ first and the emergent 
 processed spectrum is considered to be incident on the 
 other sub-shell $Sh^{BG}_{1}$. The emergent spectrum
 from the latter is the processed output of the entire
 shell $Sh_{1}$ and is used as input boundary condition
 for the next shell $Sh_{2}$. In this manner, the radiation
 field is transported outward from shell $Sh_{1}$ to $Sh_{2}$
 ... till the last shell, viz., $Sh_{N}$ is reached. This
 entire processing from shell 1, 2, ... to N, constitutes
 one iteration. Several such iterations (typically 5--10)
 are carried out
 until a set of predetermined convergence criteria are
 satisfied. The emerging spectrum from the last shell, $Sh_{N}$,
 is the desired output of the full model.
The number of shells used for a specific source is determined by
the criterion 
that the shell is optically thin
in the shortest relevant wavelength.

\subsection{Processing of transient heating of the VSG and the PAH}

    As described above, the dust components in a shell for which the
 non-equilibrium emission processes are important, are
 segregated into a separate sub-shell ($Sh^{VP}_{k}$)
 consisting only of the VSG and the 
 PAH.
 The interaction of the total incident radiation field 
 from both the surfaces of this sub-shell, ($I^{\nu}_{in}$), 
 with the VSG component, is considered using 
a code developed by us
based on the statistical
 mechanical treatment prescribed by Desert, Boulanger \& Shore (1986).
 The incident radiation, partly extinguished by the VSGs
 ($I^{\nu}_{VP} = I^{\nu}_{in} \times e^{-\tau^{\nu}_{VSG}}$),
 is considered incident on the PAH component and a similar
 computation is repeated. The final emerging spectrum 
 consists of three components: (i) the originally incident 
 radiation extinguished by {\it both} VSG as well as PAH, 
 ($I^{\nu}_{out} = I^{\nu}_{in} \times 
e^{-(\tau^{\nu}_{VSG} + \tau^{\nu}_{PAH})}$),
 (ii) the emission from the VSG component, and (iii) the
 emission from the PAH component. Whereas the first component
 is direction sensitive (the two surfaces get different
 contributions depending on the original spectrum incident 
 at the other surfaces), the latter two 
 contribute equally to the two surfaces.

   The VSG and PAH components of grains have fluctuating temperature,
mainly because their enthalpy (internal energy) is comparable to
the energy of UV or visual photons. This means, the multiphoton
absorption processes can become important (depending on the exact
radiation field and the details of thermal \& optical properties of these
grains) as they can lead to a modified temperature distribution.
An iterative method has been used here to
consider these multiphoton processes for VSGs and PAHs separately.
The method assumes a single grain in an isotropic radiation field,
and follows the evolution of the grain temperature by solving the 
relevant stochastic differential equation. 

  A scheme of between 100 to 400 levels of internal energy
(covering 0.5 eV to 200 eV) for considering discrete heating /
cooling processes; and
400 energy levels (for energies 1.25 $\times 10^{-3}$ eV to
0.5 eV) for considering the continuum processes, has been
incorporated. A total of 97 frequency grid points covering 
0.0944 $\mu$m to 5000 $\mu$m have been used. Several grid points
are densely packed around the five PAH features at 3.3, 6.2,
7.7, 8.6 and 11.3 $\mu$m.

\subsection{Radiation transport through normal grains (BG)}

 Each of the sub-shells consisting of the normal grains, $(Sh^{BG}_{k}, 
 k=1,2, ... N$), separately undergoes full radiative 
 transport calculation using the code CSDUST3 developed
 by Egan, Leung \& Spagna (1988) (see also Leung 1975). 
In CSDUST3, the moment equation of 
radiation transport and the equation
of energy balance are solved simultaneously as a two-point 
boundary value problem.
The effects
of multiple scattering, absorption and re-emission of
photons on the temperature of dust grains and the internal
radiation field have been considered self-consistently.
In addition, multi grain components, radiation 
field anisotropy and linear anisotropic 
scattering are also incorporated.

Same frequency grid of 97 points, as used for VSG and PAH, has been 
used here.
In order to avoid non-convergence problems due to sharp changes in 
optical depth at any of the frequency grids, logarithmically
increasing  radial grid spacings have been used at the inner
shell boundary. Similarly a smoothly decreasing grid 
spacings have been used near the outer shell boundary.

\subsection{The Modelling Scheme}

 The scheme aims to construct a model constrained by the observed
SED covering the entire infrared and the sub-mm / mm region.
Based on comparisons of the model predicted SEDs with the 
observed SED, various model parameters are fine tuned till
the best fit model is identified.
 The following model parameters 
are explored : (i) the total radial optical depth 
 (represented at a fiducial wavelength of 100 $\mu$m);
 (ii) exponent of the dust density distribution power law;
 (iii) the ratio of graphite component to the astronomical 
 silicate component for BGs; (iv) size of VSGs, $a_{VSG}$: either 50
 \AA \  or 10 \AA \  ; (v) size of PAH molecule / cluster, $a_{PAH}$: 
4.6 \AA \  or 8 \AA \  or 13.6 \AA \  ; (vi) relative abundance of PAH
 compared to BGs, $Y_{PAH}$ (a constant or varying with
 the radial distance) ; and
(v) the de-hydrogenation factor, $f_{de-H}$.

The inner radius ($R_{in}$), for each model
 of the spherical cloud, 
has been determined 
using the constraint that the temperature of the BG is equal to 1500 K,
the sublimation 
temperature of the normal big grains   
(graphite and astronomical silicate). The radial dust density
 distribution law has been assumed to be a power law
 $n_{d}(r) \propto r^{-{\alpha}}$, and the values of $\alpha$ that
have been explored are 0, 1 and 2. 

\section{Application of the modelling scheme}

 In order to demonstrate the usefulness of the scheme described
above, an attempt has been made to apply the same to a few H II
regions. The question is : in spite of rather simplistic treatment,
can we get any insight into physical details of these sources?

\begin{center}
\begin{table}
\caption{Input parameters of the compact H II regions}
\vskip 0.5cm
\begin{tabular}{|c|c|c|c|c|c|}
\hline
IRAS Source & $L$ & $T_{*}$ & $D_{sun}$ & $\theta_{dia}$ & $R_{out}$ \\
                 & $(L_{\odot})$ & (K)  & (kpc) & ($''$) & (pc) \\
\hline
18116--1646 & 1.6 $\times 10^{5}$ & 40,000 & 4.4 & 105 & 1.12 \\
18162--2048 & 2.8 $\times 10^{4}$ & 30,900 & 1.9 & 63 & 0.29 \\
19442+2427 & 5.4 $\times 10^{4}$ & 35,500 & 2.3 & 93 & 0.52 \\
22308+5812 & 8.8 $\times 10^{4}$ & 37,500 & 5.7 & 95 & 1.31 \\
18434--0242 & 1.0 $\times 10^{6}$ & 48,000 & 7.4 & 38 & 0.69 \\
\hline
\end{tabular}
\end{table}
\end{center}

\subsection{The sample of compact H II regions}
 
     With the advent of Infrared Space Observatory (ISO), it
 has become now possible to have precise spectroscopic
 information in the entire infrared band encompassing near to
 far infrared region. The spectroscopic results for a 
 sample of six Galactic compact H II regions, covering four of the five
 major PAH features have been published by Roelfsema {\it et al.} (1996).
 We have chosen five out of their six sources for our detailed study.
The sixth source IRAS 21190+5140,
identified with M1-78 and variously considered as H II region and 
planetary nebula (Puche {\it et al.} 1988, Acker {\it et al.} 1992),
has not been considered here.
 Although the published spectral results from 
 ISO is rather limited (6 - 12 $\mu$m), if
 the IRAS Point Source Catalog (IRAS PSC) 
measurements (at 12, 25, 60 \& 100 $\mu $m), IRAS Low Resolution Spectra
(IRAS LRS; between 8 -- 22 $\mu $m) and ground based spectroscopy around
the 3.3 $\mu $m PAH feature, are included (whenever available), 
then sufficient observational constraints can be placed on the radiative
transfer models. All these measurements have been compiled to 
construct the SEDs for the five compact H II regions, which are displayed in
Figure 2.

  The total luminosity, $L_{tot}$, 
has been taken from Roelfsema {\it et al.} (1996) for all the 
sources except for IRAS 18434-0242. The luminosity for IRAS 18434-0242
listed by them is too low compared to that estimated from the 
IRAS data. The latter has been used here for modelling.
Assuming the embedded source to be a single ZAMS star
of luminosity $L_{tot}$,
a Planckian spectral shape with corresponding temperature taken 
from Thompson (1984), has been assumed. Since distance estimates
are available for all these sources, angular sizes are required
to fix the outer radii of the clouds.
The mid infrared angular sizes are estimated by comparing the 
flux densities at 12 $\mu $m as measured by IRAS-LRS and the ISO-SWS
and the solid angle covered by the ISO-SWS.
It is assumed that the source size is smaller than the IRAS-LRS
beam and has constant brightness per unit solid angle.
The entire size of the cloud has been estimated from the 12 $\mu $m
size by using empirical relation between angular size and
the wavelength for compact H II regions (Mookerjea \& Ghosh 1999).

\begin{center}
\begin{table}
\caption{Dust parameters valid for the entire
sample of compact H II regions}
\vskip 0.5cm
\begin{tabular}{|l|l|l|}
\hline
 Dust component & Parameter (unit) & Value  \\
\hline 
 BG: & $a_{min}$ ($\mu$m) & 0.01 \\
 & $a_{max}$ ($\mu$m) & 0.25 \\
 & $\gamma $ & --3.5 \\
 && \\
 VSG: & $a_{VSG}$ (\AA \ ) & 50.0 \\
 & $Y_{VSG} $ & 4.70 $\times 10^{-4}$  \\
 && \\
 PAH: & $a_{PAH}$ (\AA \ ) & 8.0 \\
 & $Y_{PAH} $ & 4.30 $\times 10^{-3}$  \\
 & $f_{de-H} $ & 0.0 \\
\hline
\end{tabular}
\end{table}
\end{center}

\subsection{Results of modelling}

 It has been possible to get reasonable fits to the observed SEDs of
all the five compact H II regions by varying parameters
of our modelling scheme.
The predicted spectra and the observations are compared in Figure 2. 

\begin {figure*}
 \epsfysize=200.0pt
 \epsfbox {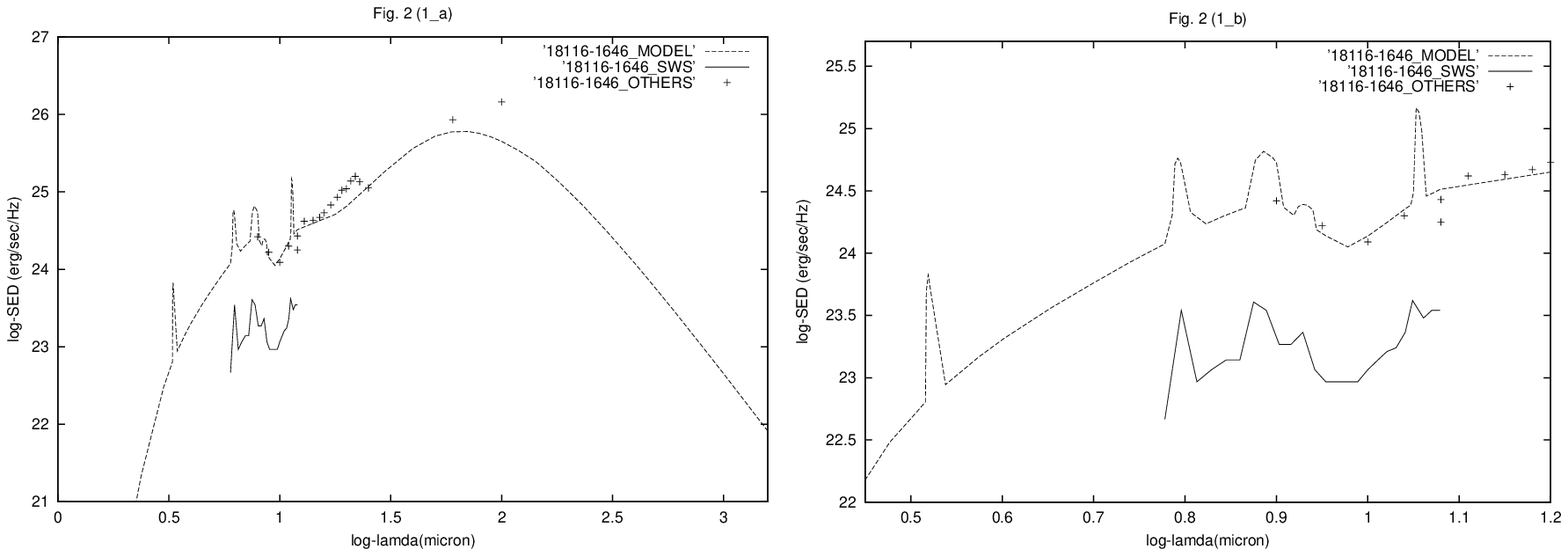}
\caption{
 Spectral energy distribution of the compact H II region
IRAS 18116-1646.
The ordinate is
the log of the flux density multiplied by the surface area of the
respective cloud. 
Solid lines
show the ISO-SWS spectra from Roelfsema {\it et al.} (1996); dotted lines show 
our best fit model spectra; diamonds show other observations. 
Other observations
include --- 3 $\mu$m  observations from de Muizon,
d'Hendecourt \& Geballe (1990), IRAS LRS spectra from Olnon \& Raimond (1986)
or from Volk \& Cohen (1989), IRAS PSC flux densities,  sub-mm observations 
from McCutcheon {\it et al.} (1995), Jenness, Scott, \& Padman (1995) or 
Barsony (1989), and 1.3 mm observation
of Chini, Krugel, \& Kreysa (1986). In order to show the PAH features clearly, 
mid IR region of the SEDs are shown separately (on the right).
}
\end {figure*}
\setcounter{figure}{1}
\begin {figure*}
 \epsfysize=200.0pt
 \epsfbox {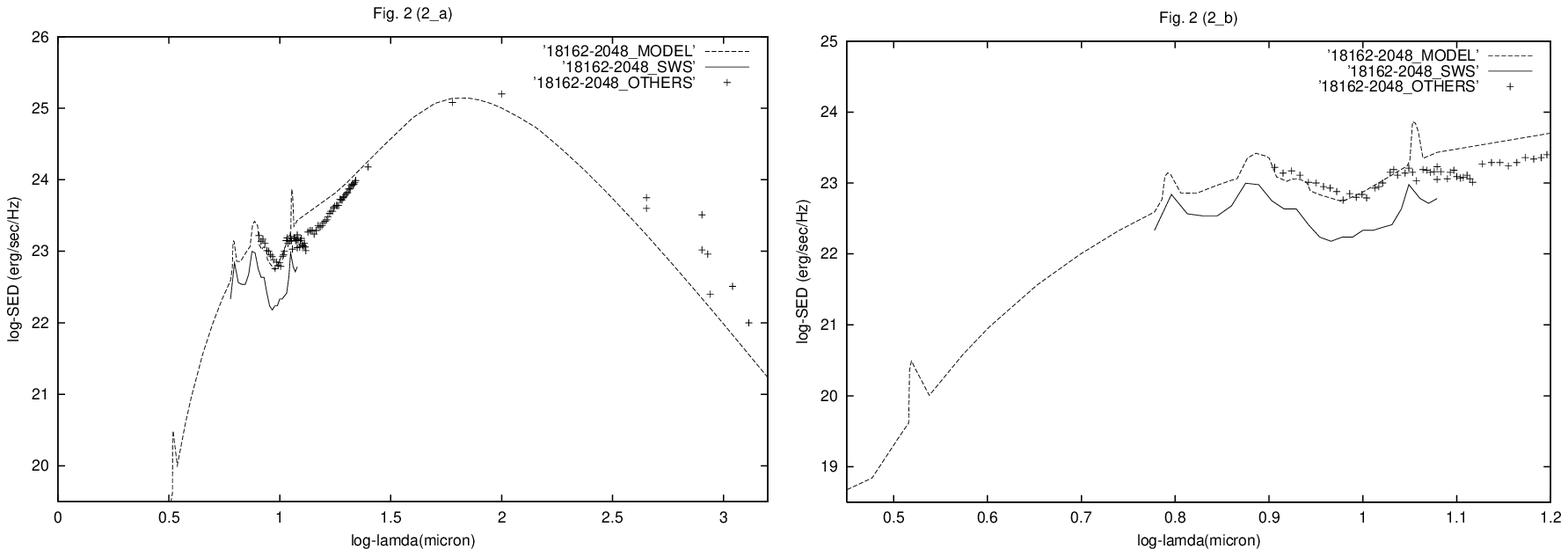}
\caption{
Continued ... (for IRAS 18162-2048)
}
\end {figure*}
\setcounter{figure}{1}
\begin {figure*}
 \epsfysize=200.0pt
 \epsfbox {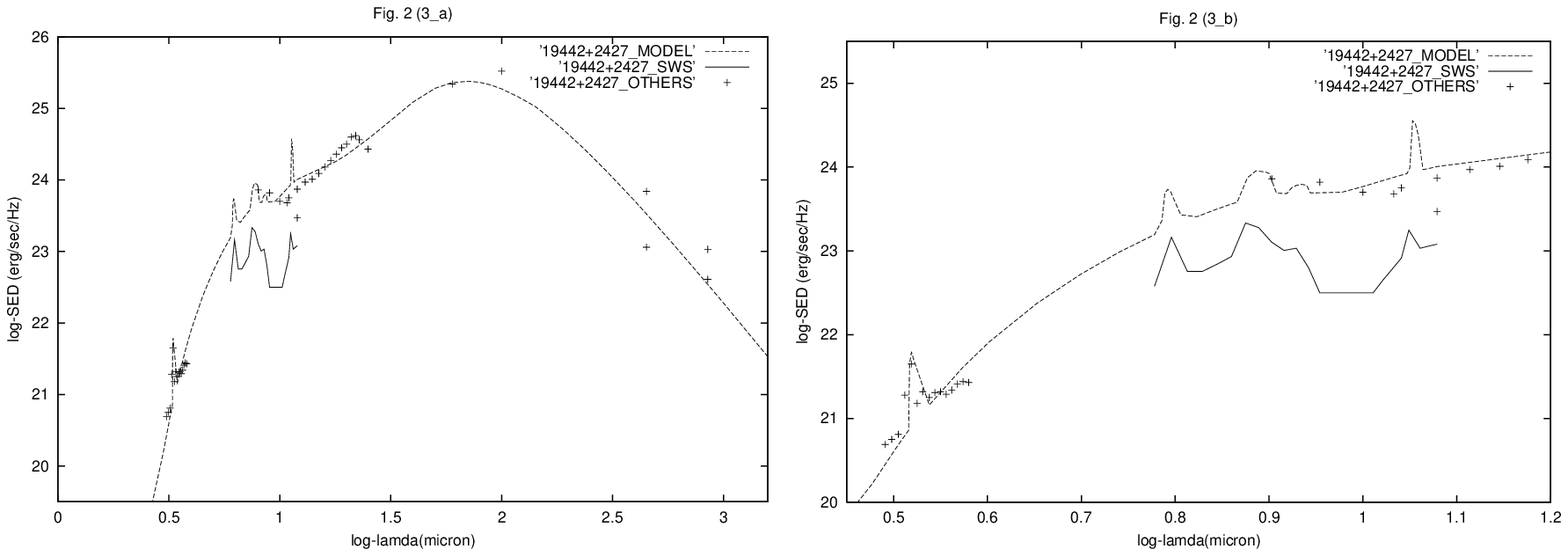}
\caption{
Continued ... (for IRAS 19442+2427)
}
\end {figure*}
\setcounter{figure}{1}
\begin {figure*}
 \epsfysize=200.0pt
 \epsfbox {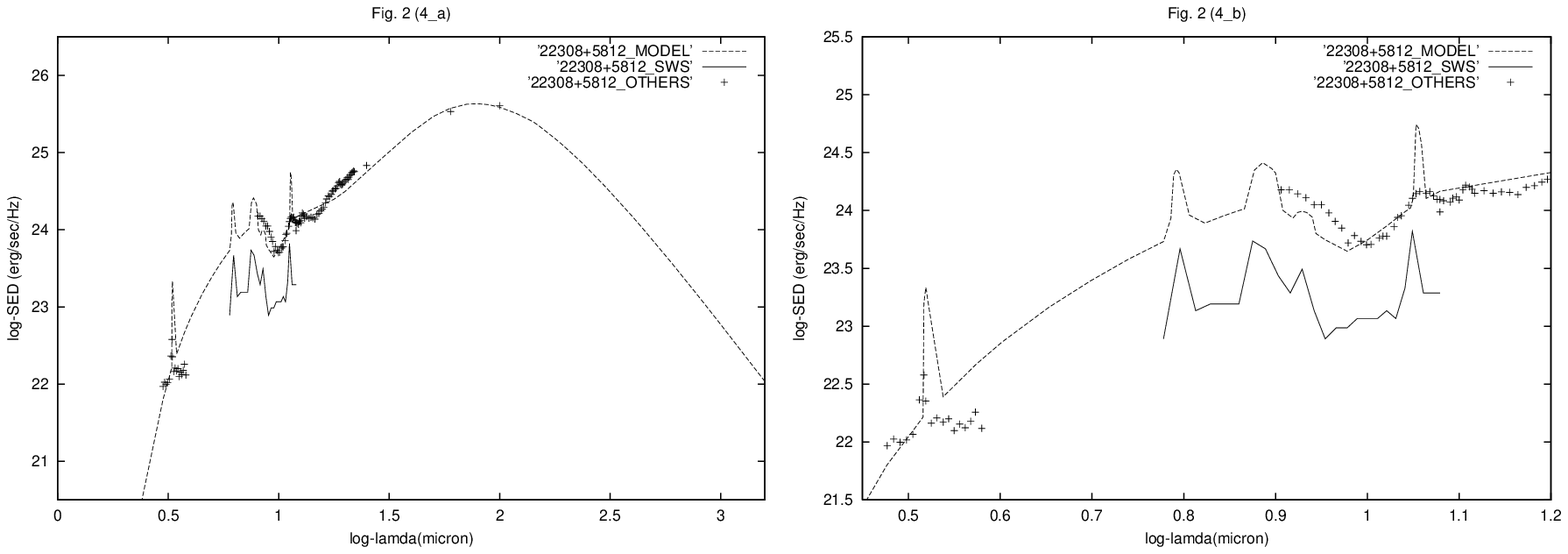}
\caption{
Continued ... (for IRAS 22308+5812)
}
\end {figure*}
\setcounter{figure}{1}
\begin {figure*}
 \epsfysize=200.0pt
 \epsfbox {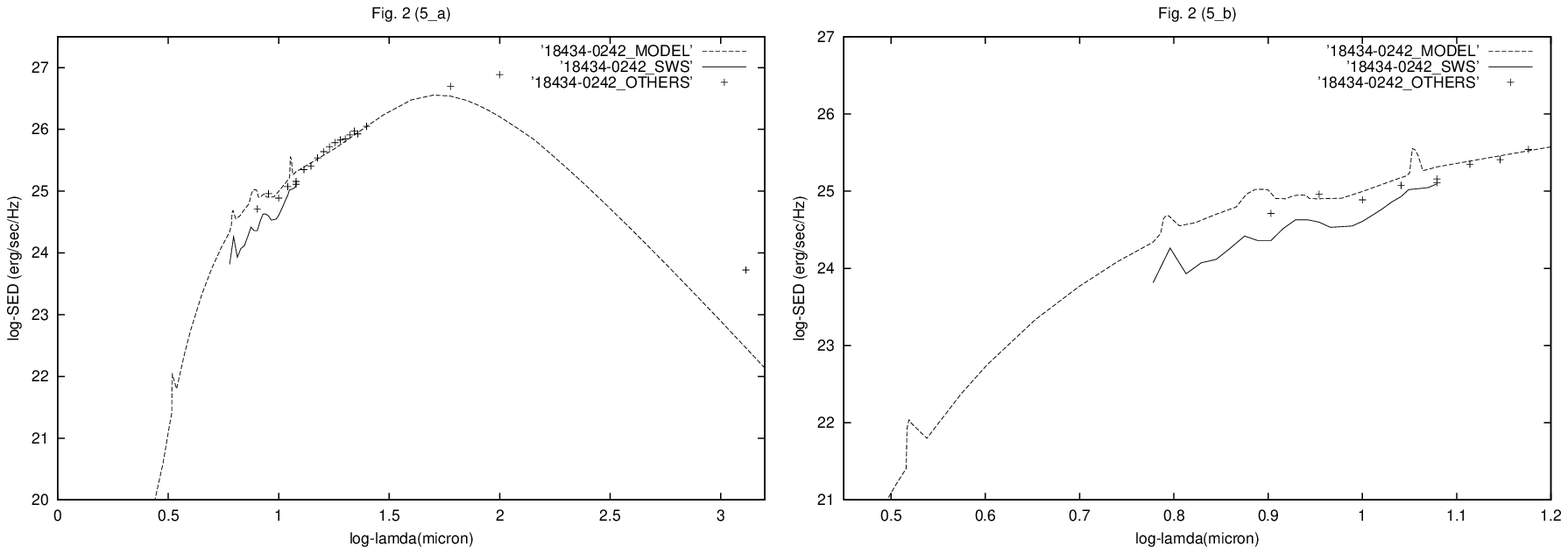}
\caption{
Continued ... (for IRAS 18434-0242)
}
\end {figure*}
 
 The following comments are valid for all the five compact H II 
regions studied here.
 It was found that the models with uniform density
distribution i.e. $n(r) \propto r^{0}$ (as opposed 
to $n(r) \propto r^{-1}$ or $r^{-2}$) gave much better fits
 to the SEDs.
The VSGs with $a_{VSG}$ = 50 \AA \ and the PAHs with intermediate
size (i.e. $a_{PAH}$ = 8 \AA \ ) give better fits to the respective
spectra. The de-hydrogenation factor, $f_{de-H}$, needs to be
zero (corresponding to ${\rm N_{H}=\sqrt {6\times N_{C}}}$)
in order to fit the relative strengths of the PAH 
features for all the five sources. 
This value of $f_{de-H}$, is typical for the types of PAH 
which have been strongly proposed in the literature, viz.,
Coronene \& Ovalene (Leger \& Puget, 1984). 
In addition, this $f_{de-H}$ is consistent with the value of
$a_{PAH}$ (8 \AA \ ) inferred from our modelling, since such PAHs are
expected to be completely hydrogenated in the emission zones
(Allamandola, Tielens \& Barkar, 1989).
Table 2 lists those best fit
parameters related to the various dust components, 
which are valid for the entire sample of compact H II 
regions. 

It has been found that, whereas the BG  and VSG components
should exist throughout the cloud, it is absolutely necessary
that the PAH component must be confined
to a thin inner region ($R^{PAH}_{out} << R_{out}$), in order
to reproduce the PAH features. The size of this region is quantified by
a parameter $\eta_{PAH}$, which is defined as :
$\eta_{PAH} = ((R^{PAH}_{out} - R_{in})/(R_{out} - R_{in}))$.
This parameter had to be varied for each source, 
till a good fit to the spectrum was obtained.
In addition, the abundance of PAH relative to BGs, $Y_{PAH}$,
needed to be increased by a factor 10 relative to the normal
value obtained by Desert, Boulanger \& Puget (1990). However this does
not lead to any conflict with the available carbon, since 
$\eta_{PAH} << 1$.
 The values of the best fit parameters specific to 
each source, are presented in Table 3.

\hoffset -2.5cm
\begin{center}
\begin{table}
\caption{Best fit parameters of the compact H II regions as determined by modelling}
\vskip 0.5cm
\begin{tabular}{|c|c|c|c|c|c|c|c|}
\hline
IRAS Source Name & $R_{in}$ & $R_{out}$ & $n_{H}$ & $M_{Tot} $ & $\tau^{tot}_{100}$ & Graphite : Silicate& $\eta_{PAH}$ \\
                 & (pc) & (pc) & $(cm^{-3})$ & ($M_{\odot}$) & & (\% : \%) &  \\
\hline 
 &&&&&&& \\
18116--1646 & 1.4 $\times 10^{-3}$ & 1.12 &$1.32 \times 10^{4}$ & $1.9 \times 10^{3}$ & 0.056 & 75 : 25 & 2.1 $\times 10^{-2}$ \\
18162--2048 & 5.3 $\times 10^{-4}$ & 0.29 & $1.32 \times 10^{5}$ & $3.3 \times 10^{2} $ &0.14 & 88 : 12 & 2.7 $\times 10^{-2}$ \\
19442+2427 & 7.5 $\times 10^{-4}$ & 0.52 &$5.30 \times 10^{4}$ & $7.7 \times 10^{2} $ & 0.10 & 95 : 5 & 1.5 $\times 10^{-2}$ \\
22308+5812 & 1.5 $\times 10^{-3}$ & 1.31 &$1.38 \times 10^{4}$ & $3.2 \times 10^{3} $ & 0.068 & 77 : 23 & 1.3 $\times 10^{-2}$ \\
18434--0242 & 2.2 $\times 10^{-3}$ & 0.69 &$5.30 \times 10^{4}$ & $1.8 \times 10^{3} $ & 0.14 & 95 : 5 & 5.5 $\times 10^{-2}$ \\ 
 &&&&&&& \\
\hline
\end{tabular}
\end{table}
\end{center}

 \section{Discussion }

 The following inferences can be drawn about the sources modelled
here, provided the basic assumptions (e.g. spherical symmetry;
sources of energy located only at the centre of the cloud; etc)
are not at great variance from the reality.
 
  The most
favoured radial dust density distribution law,
{\it for all five sources}, turns out 
to be of uniform density. This can perhaps be understood in terms of 
the far infrared constraints (IRAS-PSC 60 \& 100 $\mu$m data).
If the dust density is falling with radial distance, 
then in order to fit the FIR part of the SED, so high a
dust density is required at the vicinity of the embedded
ZAMS star, that the mid infrared emission becomes invisible.
This problem can perhaps be avoided in a non-spherically
symmetric geometry. We have explored the effects of
relaxing the assumption that  $R_{in}=R_{0}$,
where $R_{0}$ is the radial distance at which (BG) grain temperature
becomes equal to the sublimation temperature (1500 K).
The most important effect of making 
$R_{in} >  R_{0}$,
is to drastically modify the near and mid infrared
continuum level of the predicted spectrum. In addition, the
role of non-equilibrium processes vis-a-vis thermal 
equilibrium emission of the PAH features, as well as the
continuum due to VSG, changes significantly.

   A quick perusal of Figure 2 and Tables 2 and 3 
brings out the following facts:

1) All the compact H II regions considered here, are deeply embedded stars;
total optical depth at 100 $\mu$m in the range of 0.056 -- 0.14. This is
necessary to explain the far IR spectra observed by IRAS.   

2) PAH is confined only to a thin central shell;
the thickness of this shell
being just a few percent (1.3 -- 5.5 \%) of the total 
thickness  of the dust cloud. 
As these sources are optically thick at mid
IR, if the PAH is distributed throughout the cloud, its emission which
occurs in the inner hot region where high energy photons responsible for
non-equilibrium processes are present, will be absorbed by the outer
cooler shells, and PAH features will not be detectable.

3) The BGs are dominated by graphites, with silicates contributing less
than 25 \%.
The latter has been tied down rather precisely by the 10 $\mu$m silicate
feature.

4) ISO-SWS fluxes are generally much smaller than IRAS fluxes at similar
wavelengths, indicating that SWS is not sampling full emission at mid IR
and the source sizes at these wavelengths are much larger 
than the SWS beam size 
($14'' \times 20''$).
With this in mind, we have not tried to fit the absolute fluxes of the SWS
but only used its shape as indicative of the importance of PAH molecules.

    Following comments can be made about the individual sources:

\begin{itemize}

\item {\bf IRAS 18116--1646} :  It has relatively lower optical
depth. The fit to the IR data is quite reasonable, except at 100 $\mu$m
where IRAS flux is higher; no sub-mm observation exists for this source.

\item {\bf IRAS 18162--2048} : This source (GGD27) was originally thought
to be a HH object. However now it has been established as a star forming
region with reflection nebulosity as well as outflow (see for example
Stecklum {\it et al.} 1997). The region has several near IR and mid IR sources;
the source of energy being close to IRS2. The size of this source at 
sub-mm wavelengths is $\sim 1'$ (McCutcheon {\it et al.} 1995), 
consistent with the
size for the best fit model. The mass of the envelope estimated by our
model is not far from the estimate of Yamashita {\it et al.} (1987), viz., 200$
M_{\odot}$. They have proposed a disk geometry for this source. This source
has very high optical depth. The fit to the IR data is quite reasonable
but at the sub-mm wavelengths the calculated flux densities are lower than
the observed ones.

\item {\bf IRAS 19442+2427} :  This source lies in the H II region S87. The
size of this source at sub-mm wavelengths is $\sim 1'$ (Jenness, Scott and
Padman, 1995), consistent with the size for the best fit model. This
source has medium optical depth. The fit to all the observations from 3
$\mu $m to 850 $\mu$m is quite reasonable.

\item {\bf IRAS 22308+5812} :  It has relatively lower optical
depth. The fit to the IR data is quite reasonable; no sub-mm observation
exists for this source.

\item {\bf IRAS 18434--0242} : This source is the most luminous source with
high optical depth.  There are no IR observations for this source
other than those from IRAS and ISO. IRAS PSC 100 $\mu$m as well as 
1.3 mm observations 
are higher than calculated.

\end{itemize}

  From the above, we conclude that our new scheme of radiative
transfer which includes non-equlibrium processes (transient heating
of the grains / PAH / VSG) in addition to the emission in thermal equilibrium,
can give important physical insight into Galactic star forming regions.
If the simplifying geometrical assumptions of our scheme are valid, 
then several important inferences can be made about the five compact H II
regions considered for modelling here.

\centerline{\bf Acknowledgements}
\vskip 0.5cm
It is a pleasure to thank Bhaswati Mookerjea for her help in
preparing some input parameters.

\label{lastpage}
\end{document}